%% file: main.tex
\documentclass[letterpaper,twocolumn,10pt]{article}
\usepackage{style/usenix}

\usepackage{style/wspr}
\usepackage{enumitem}
\usepackage{graphicx}
\usepackage{float}
\usepackage{booktabs}
\usepackage{xurl}
\usepackage{minted} 

\usepackage[dvipsnames]{xcolor}
\usepackage{mdframed}
\usepackage{listings}
\usepackage{adjustbox}
\usepackage{subfig}
\usepackage{diagbox}

\begin{document}

\newcommand{\etal}{\textit{et al.}\xspace}

\newcommand*\optionion[1]{
  \color{red}{#1}\color{black}
}
\newlist{questions}{enumerate}{2}
\setlist[questions,1]{label=RQ\arabic*.,ref=RQ\arabic*}
\setlist[questions,2]{label=(\alph*),ref=\thequestionsi(\alph*)}
\setlist[questions]{topsep=0pt,itemsep=-1ex,partopsep=1ex,parsep=1ex}

\newlist{contributions}{enumerate}{2}
\setlist[contributions,1]{label=C\arabic*.,ref=C\arabic*}
\setlist[contributions,2]{label=(\alph*),ref=\thecontributionsi(\alph*)}
\setlist[contributions]{topsep=0pt,itemsep=-1ex,partopsep=1ex,parsep=1ex}

\title{Characterizing phishing pages by JavaScript Capabilities}

\input{macros/helpers}

\input{macros/numbers}

\input{tex/authors}

\maketitle

\input{tex/abstract.tex}
\input{tex/introduction.tex}

\input{tex/background.tex}
\input{tex/methods.tex}
\input{tex/data.tex}

\input{tex/results.tex}
\input{tex/discussion.tex}
\input{tex/related_work.tex}
\input{tex/limitations_future_work.tex}
\input{tex/conclusion.tex}

\bibliographystyle{plain}
\bibliography{bib/rel_work.bib, bib/relwork.bib}
\input{tex/appendix.tex}

\end{document}

%% file: macros/helpers.tex
\definecolor{aoenglish}{rgb}{0.0, 0.5, 0.0}

\newif\ifsubmit\submittrue
\ifsubmit
\newcommand{\textcomment}[1]{}
\newcommand{\alex}[1]{}
\else
\newcommand{\textcomment}[1]{\noindent\textcolor{aoenglish}{\textit{\textbf{Comment:} #1 \\}}}
\newcommand{\alex}[1]{\textcolor{purple}{\textbf{Alex:} #1}}
\fi

\ifsubmit
\newcommand{\RC}[1]{}
\else
\newcommand{\RC}[1]{[\textcolor{blue}{RC~#1}]}
\fi

\newcommand\encircle[1]{\microtypesetup{disable}\tikz[baseline=(X.base)] \node (X) [draw, shape=circle, inner sep=0em,text width=1em, text centered] {\scriptsize #1};\microtypesetup{enable}}

\newcommand{\red}[1]{\textcolor{red}{#1}}

\newcommand{\code}[1]{\mintinline{javascript}{#1}}
\newcommand{\greencode}[1]{\textcolor{aoenglish}{\code{#1}}}

\newcommand{\tool}{{\sc Argus}\xspace}
\newcommand{\systemname}{{\sc Argus}\xspace}
\newcommand{\toolurl}{\url{https://secureci.org/argus}}
\newcommand{\pocurl}{\url{https://secureci.org/poc}\xspace}
\newcommand{\vwurl}{\url{https://poc.secureci.org}\xspace}
\newcommand{\benchmarkname}{VWBench}
\newcommand{\gh}{GitHub}
\newcommand{\js}{JavaScript}
\newcommand{\kp}{\emph{KitPhishr}}
\newcommand{\ji}{Jaccard index}
\newcommand{\gci}{GitHub CI}
\newcommand{\ghasttool}{{\sc Ghast}\xspace}
\newcommand{\githubtool}{{\sc GitSec}\xspace}
\newcommand{\gwchecker}{{\sc Gwchecker}\xspace}
\newcommand{\actions}{\code{Actions}}
\newcommand{\action}{\code{Action}}
\newcommand{\codeql}{{\sc CodeQL}\xspace}
\newcommand{\vwbench}{{\sc VWBench}\xspace}
\newcommand{\plugins}{Plugins}

\newcommand{\tbl}[1]{Table~\ref{#1}}
\newcommand{\sect}[1]{Section~\ref{#1}}
\newcommand{\fig}[1]{Figure~\ref{#1}}
\newcommand{\apdx}[1]{Appendix~\ref{#1}}
\newcommand{\lst}[1]{Listing~\ref{#1}}
\newcommand{\algo}[1]{Algorithm~\ref{#1}}
\newcommand{\exploitrepo}{\url{https://anonymous.4open.science/r/poc-0B60/}\xspace}

\definecolor{actioninputcolor}{rgb}{1.0, 0.49, 0.0}
\newcommand{\inputsourceicon}{\textcolor{actioninputcolor}{\faBolt}}
\newcommand{\taintsourceicon}{\textcolor{red}{\faUserSecret}}
\newcommand{\taintsinkicon}{\textcolor{red}{\faBomb}}
\newcommand{\usesvulnerableaction}{\textcolor{red}{\faUnlink}}
\newcommand{\flowsicon}{\faMailReply}
\newcommand{\sanitizeicon}{\textcolor{green}{\faLock}}
\newcommand{\hoticon}{\faFire}

\newcommand{\DraftToDo}[1]{%
\begin{mdframed}[backgroundcolor=red!10,linecolor=red!50]
\textbf{TODO:} #1
\end{mdframed}
}

\newcommand{\DraftNote}[1]{%
\begin{mdframed}[backgroundcolor=RoyalPurple!20,linecolor=RoyalPurple!50]
\textbf{Note for Kap:} #1
\end{mdframed}
}

\newcommand{\KapSays}[1]{%
\begin{mdframed}[backgroundcolor=Green!20,linecolor=RoyalPurple!50]
\textbf{Note from Kap:} #1
\end{mdframed}
}

\newcommand{\BradSays}[1]{%
\begin{mdframed}[backgroundcolor=Blue!20,linecolor=RoyalPurple!50]
\textbf{Note from Brad:} #1
\end{mdframed}
}
\lstdefinelanguage{JavaScript}{
  keywords={typeof, new, true, false, catch, function, return, null, catch, switch, var, if, in, while, do, else, case, break},
  keywordstyle=\color{blue}\bfseries,
  ndkeywords={class, export, boolean, throw, implements, import, this},
  ndkeywordstyle=\color{darkgray}\bfseries,
  identifierstyle=\color{black},
  sensitive=false,
  comment=[l]{//},
  morecomment=[s]{/*}{*/},
  commentstyle=\color{purple}\ttfamily,
  stringstyle=\color{red}\ttfamily,
  morestring=[b]',
  morestring=[b]"
}

\lstdefinestyle{jsstyle}{
  language=JavaScript,
  frame=single,
  basicstyle=\ttfamily\footnotesize,
  keywordstyle=\color{blue},
  stringstyle=\color{teal},
  commentstyle=\color{gray}\ttfamily\itshape,
  numbers=none,
  breaklines=true,
  columns=flexible,
  showstringspaces=false,
  captionpos=b
}

\lstset{
   language=JavaScript,
   extendedchars=true,
   basicstyle=\footnotesize\ttfamily,
   showstringspaces=false,
   showspaces=false,
   numbers=left,
   numberstyle=\footnotesize,
   numbersep=9pt,
   tabsize=2,
   breaklines=true,
   showtabs=false,
   captionpos=b
}

%% file: macros/numbers.tex
\newcommand{\totalPages}{1,328,917} 
\newcommand{\totalPhishingDomains}{XXXX}
\newcommand{\totalPageWithJavascript}{YYYYY}
\newcommand{\totalPagesWithJavascriptFP}{952,155}
\newcommand{\totalPagesClusterable}{434,495}
\newcommand{\totalClusters}{9,306}
\newcommand{\TotalAPIs}{17,753}
\newcommand{\APIcutoff}{8}
\newcommand{\daysCrawled}{523}
\newcommand{\clientcheckPages}{19,869}
\newcommand{\clientcheckCluster}{504}
\newcommand{\clientcheckTopClusterPrecent}{23\%}
\newcommand{\totalZipFiles}{7,273}
\newcommand{\kitsToReview}{5,871}
\newcommand{\totalUniqueZips}{3,713}
\newcommand{\totalNumberOfKits}{2,262}
\newcommand{\totalNumberOfKitFamilies}{885}
\newcommand{\totalKitPages}{4,180}
\newcommand{\totalKitClusters}{764}
\newcommand{\gtFMI}{0.98}
\newcommand{\gtVS}{0.92}
\newcommand{\gtcompleteness}{0.92}
\newcommand{\gthomogeneity}{0.92}
\newcommand{\dbscanEpsi}{0.05}

\newcommand{\badClusters}{360}
\newcommand{\badClustersPercentage}{2.52}
\newcommand{\badClustersPages}{37,774}
\newcommand{\badClustersPagesPercentage}{8.00}
\newcommand{\numReviewers}{5}
\newcommand{\pairsReviews}{\emph{5,871}}

%% file: tex/authors.tex
\author{
{\rm Aleksandr Nahapetyan}\\
North Carolina State University\\
\textit{anahape@ncsu.edu}
\and
{\rm Kanv Khare}\\
North Carolina State University\\
\textit{kkhare@ncsu.edu}
\and
{\rm Kevin Schwarz}\\
North Carolina State University\\
\textit{kmschwa3@ncsu.edu}
\and
{\rm Bradley Reaves}\\
North Carolina State University\\
\textit{bgreaves@ncsu.edu}
\and
{\rm Alexandros Kapravelos}\\
North Carolina State University\\
\textit{akaprav@ncsu.edu}
} %

%% file: tex/abstract.tex
\begin{abstract}
Phishers achieve large-scale attacks by using ready-to-deploy phishing websites (phishing kits) to rapidly launch campaigns that leverage specific data exfiltration, evasion, or mimicry techniques.
In contrast, researchers and defenders continue to rely on manual analysis to identify features for kit fingerprinting.
In this paper, we examine the link between a page's client-side behavior and the underlying phishing kit used, enabling automated aggregation of phishing pages.
Our key insight is that client-side techniques make heavy use of browser APIs, which, in turn, differentiate underlying kits based on their feature sets. 
Using an instrumented browser and a URL fuzzing utility, we collected traces from 1,328,917 pages and recovered kit archives for \totalKitPages{} pages between August 2023 and January 2025.
For the labeled subset, we find that clustering based on the set of browser APIs executed yields 98\% accuracy in grouping them by the underlying kit. We also find that 434,495 phishing pages execute enough browser APIs to cluster into 9,306 clusters, compressing multi-lingual phishing pages across various domains into a single cluster.
Our findings show that analysts and researchers can leverage the complexity of client-side phishing code to track phishers' kit deployments in the wild.
\end{abstract}

%% file: tex/introduction.tex
\section{Introduction}
\label{sec:intro}

Web-based phishing attacks, where a webpage mimics an official entity or creates a sense of urgency to trick the user into submitting personal information or granting access to their machine, have been increasing over the last 5 years~\cite{apwg,CISA}. Phishing attacks can have high-profile targets, like the Non-Government Organizations (NGOs) and government workers targeted in 2021~\cite{intelligenceNewSophisticatedEmailbased2021,SophisticatedSpearphishingCampaign} and lead to more sophisticated cyberattacks and data breaches~\cite{ThreatActorLeverages2024}. Phishing kits, ready-to-deploy software packages sold at illicit markets for launching phishing attacks, have lowered the entry barrier for malicious actors. Sellers often market phishing kits as bundles of quality-of-life features for attackers, such as built-in evasions from automated crawlers, exfiltration to Telegram channels, and obfuscation~\cite{tejaswiLeakyKitsIncreased2022,oest_inside_2018}. Deploying these kits can be as easy as uploading them to free hosting providers and mass-sending multiple links that exfiltrate credentials to an endpoint controlled by the attacker. As phishing kits receive software updates from the original developer or the phishers who bought them, they split into variations belonging to the same ``phishing kit family''.

One of the selling points of phishing kits is their ability to evade researchers and analysts, extending the time between deployment and discovery and increasing the number of victims who visit the page without a browser warning. While server-side logic of kits can only make assumptions about the system based on the IP address and user agent, client-side JavaScript code, through API calls to the browser, can query the user's system for CPU core counts, detect memory overhead, request user interaction, or call out to a third-party bot detection like CloudFlare\footnote{Similar to the phishing page that stole credentials from Troy Hunt, the maintainer of HaveIBeenPwned~\cite{SneakyPhishJust2025}}~\cite{zhangCrawlPhishLargescaleAnalysis2021,zhangImSPARTACUSNo2022}. In the end, the JavaScript logic in a kit can range from a simple user-agent-based redirection to an AES-encrypted script that dynamically decrypts itself, identifies the browser through a series of API calls, and renders the page after confirming the victim is using a mobile device. Web pages rely on browser APIs to facilitate this behavior, as it requires functionality that is not exposed to the loaded page.

This paper aims to aid researchers and analysts by automatically differentiating groups of phishing pages based on the pages' JavaScript behaviors. This automates a previously manual process and enables us to measure the popularity of different client-side techniques across these groups. Figure~\ref{fig:pages_clusters_per_month} shows the relative difference in volume of phishing domains compared to the number of clusters monthly active.
We focus on the following 10 techniques used by phishing pages: fingerprint exfiltration, client-side IP checks, timing-based bot detection, encoding-based obfuscation, encryption, dynamic script execution, basic fingerprinting, dynamic script injection, CloudFlare Turnstile, and pop-ups. These techniques harvest credentials~\cite{sanchez-rolaRodsLaserBeams2023}, evade crawlers~\cite{zhangCrawlPhishLargescaleAnalysis2021,oest_phishfarm_2019}, or obfuscate code to avoid analysis~\cite{jsobf-imc20,fv8-sec24}, all of which are sought-after capabilities of phishing kits~\cite{oest_inside_2018}.

In 17 months, we collected browser API traces from 1.3 million pages and successfully identified the deployed kit behind \totalKitPages{} of them. On our labeled dataset, we find that clustering over the set of browser APIs executed yields an Fowlkes-Mallows Index (FMI) of \gtFMI{} (73\% with a rebalanced ground truth) and a validity score of \gtVS{}. We then apply the clustering approach to \totalPagesClusterable{} pages, each of which executes at least 8 distinct browser APIs, grouping them into \totalClusters{} clusters. We find that among these clusters, fingerprinting and UI event handlers are near-universal techniques.
In addition to prior techniques examined by prior work, we examine two emerging techniques: client-side IP reputation checks and CloudFlare Turnstile verification, among a small set of clusters, finding that while sophisticated threat actors have cited these techniques, they are rare on phishing feeds. Finally, we evaluate the frequently used obfuscation techniques by these clusters, finding that kits favor `eval' statements and base64 encoding, with few instances of AES encryption and dynamic script generation. 

\noindent{Overall, we offer the following contributions:}

\begin{contributions}
    \item We find that \emph{browser API usage alone} is sufficient to isolate and distinguish known phishing kit families. With a ground truth dataset of \totalNumberOfKitFamilies{} kit families deployed on \totalKitPages{} URLs, we achieve accuracy metrics of a \gtFMI{} Fowlkes-Mallows score (0.73 when rebalanced for a single instance of a mass-deployed kit) and an \gtVS{} V-measure in the clusters of these pages relative to their kits used.
    \item We experimentally show that browser APIs common on the web (DOM, SVG, and CSS APIs), as well as property accesses, serve as a valuable signal for identifying kits, as they indicate the kit developer's choices, and that the more sophisticated a page's client-side logic is, the more indicative it is of the underlying kit.
    \item On the unlabeled dataset, by propagating techniques from member pages to the cluster as a whole, we find that UI-interactivity and basic fingerprinting are near-universal in the ecosystem. At the same time, mouse detection via browser APIs, CloudFlare Turnstile embedding, and dynamic script creation are still relatively rare. 
\end{contributions}

This work stands apart from prior research by automatically differentiating pages through dynamic rather than static analysis. As these features are tied to the capabilities a prospective kit buyer is looking for, they are less likely to vary across deployments of the same kit family. Compared to the prior attempt at kit identification~\cite{castanoPhiKitAPhishingKit2023}, of $F_1=9.03\%$ and $F_1=31.11\%$ with DOM clustering and URL path-based signatures, browser API usage yields more accurate grouping compared to the ground truth.

\input{figures/clusters_per_month}

%% file: figures/clusters_per_month.tex
\begin{figure}[t]
  \includegraphics[width=\columnwidth]{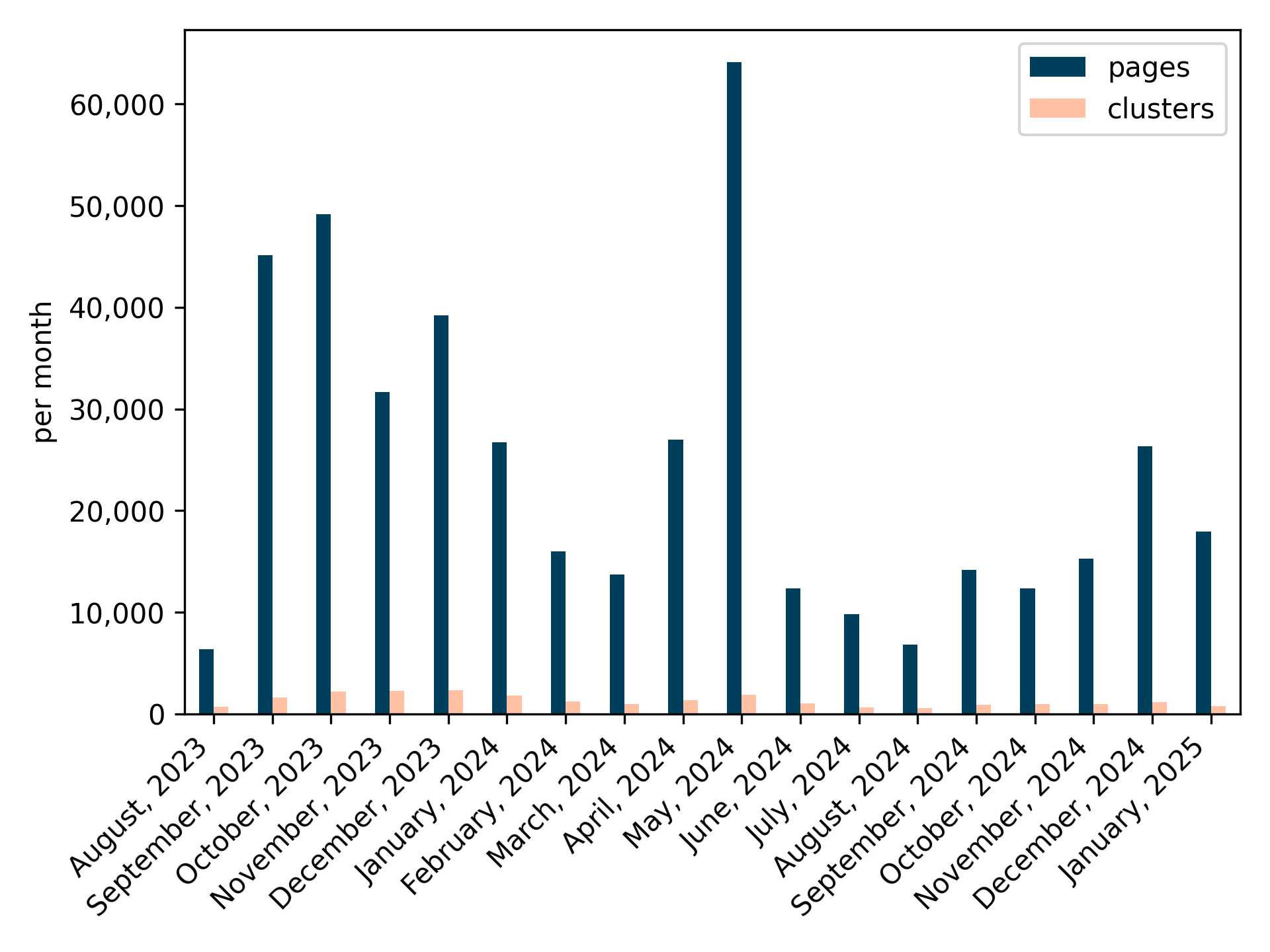}
  \caption{Comparison between monthly unique domains observed that execute 8 distinct browser APIs vs the monthly clusters observed based on dynamic behaviors. We see a drastic, non-linear reduction of phenomena that need to be investigated monthly.}
  \label{fig:pages_clusters_per_month}
\end{figure}

%% file: tex/background.tex
\section{Background}
\label{sec:background}

In this section, we provide a background on current developments in phishing as a phenomenon and adversarial JavaScript techniques, the parties involved in the phishing ecosystem, and the clustering techniques used in this work.

\subsection{The Phenomenon of Phishing}
Phishing is a form of social engineering where an adversary pretends to be a trusted entity to steal a user's credentials or gain access to a specific machine, network, or account to which the user has access. While delivery mechanisms vary, most phishing eventually leads to a webpage that requests some personal information (usually credentials) from the user. 
Because some legitimate websites use web fingerprints as a secondary authentication vector, phishers now also use browser fingerprinting APIs to identify real users and exfiltrate fingerprints to pair with stolen credentials~\cite{sanchez-rolaRodsLaserBeams2023}. 

Phishing is ever-evolving and still growing in prevalence. Groups that track phishing saw an increase in the number of phishing domains in the 2020s. The most popular target sectors vary each year, but include software-as-a-service and webmail services (Q3 2020), financial services (Q3 2021), and social media (2024)~\cite{APWGPhishingActivity}. In Q3 2024 alone, the Anti-Phishing Working Group (APWG) reported 900,000 phishing attacks. 

As more enterprises and researchers study and combat phishing, phishers respond with new countermeasures to prevent automated crawling and phishing detection, collectively called ``cloaking''~\cite{zhangCrawlPhishLargescaleAnalysis2021}.
If a phishing page determines the client is not a viable victim (e.g., a crawling bot, not in a specific country, etc.), it takes actions not to serve real phishing content.
The page may halt with an empty DOM or redirect to a benign page, a long-dead phishing page, or an affiliate marketing page.

Cloaking techniques are broadly classified into server-side and client-side techniques~\cite{oest_inside_2018}.
Server-side techniques are stealthier, but rely on limited information about the client, resorting to precompiled blocklists or allowlists of IP addresses, user agents, or HTTP referers.

Client-side techniques allow for richer evasion strategies but are also more detectable. 
Phishing pages use browser APIs to trigger permission pop-ups to identify crawling browsers, which often cannot interact with the whole browser UI. 
Because cloaking is technically very similar to legitimate bot-detection and abuse prevention, 
phishers use CAPTCHA and click-through pages as client-side cloaking. Recently, phishing pages have used Cloudflare Turnstile, a popular widget for identifying automated browsers. 

Even the URL features of a phishing link contain techniques that have evolved to respond to anti-phishing research. Phishing pages frequently use URL shorteners (public or private) to obfuscate the final destination, landing pages requiring a user to follow hyperlinks to the actual page, and free web hosting with trustworthy top-level domains (TLDs). 

\subsection{Phishing as an ecosystem}
Phishing is a logistical and technical endeavor as a phisher must develop an effective phishing page with cloaking, acquire a robust hosting for it, entice a victim to visit it via SMS, email, or social media, exfiltrate the phished data, and then monetize the stolen credentials. This technical and logistical complexity, combined with interest from potential phishers, has created an underground economy to facilitate each step.

Phishing facilitators sell bundles of customizable or ready-to-deploy phishing pages known as `phishing kits.'
They vary in features, sophistication, and cost. 
Combined with phishing-as-a-service providers, who offer phishing kits, hosting services, and continued support, lowers the barriers to entry into the ecosystem. Prior work has shown that phishing kits may steal credentials from their customers' kit deployments ~\cite{covaThereNoFree,mccalleyAnalysisBackDooredPhishing2011}, adapt or `borrow' features from other kits~\cite{intelligenceFrankenphishTodayZooBuilt2021}, and sometimes tie to specific actors ~\cite{unit42ThreatActorGroups2024}. In this paper, when two kits share the same set of features and differ only in minor additions to new IPs in the blocklists or in different directory structures, we refer to them as the same `kit family.'
Phishing systems may store credentials on the same server, risking loss when the page is inevitably taken down, but a more common practice is to send them via instant messaging.

The credential sales part of the ecosystem, which simplifies monetization, has also adapted to modern MFA/2FA practices. With prior work showing that a browser fingerprint is enough to trick online services into triggering an MFA bypass~\cite{linPhishSheepClothing2022}, and has an increasing effect on the costs of stolen credentials~\cite {sanchez-rolaRodsLaserBeams2023}.

\subsection{JavaScript}
Originally meant as a way of adding interactivity to web-
pages; the \js{} ecosystem has evolved to allow varying low-level features to webpages through Browser APIs. HTML DOM APIs enable developers to modify page appearance, while LocalStorage and IndexDB provide write access to the browser's internal storage buckets; meanwhile, the File System API allows access to the user's real machine's storage. Browser APIs can be function calls (or constructors), property reads, and property writes. Most of the privileged functionality comes from function calls.

The dynamic nature of JavaScript enables a variety of techniques for concealing itself from analysis and detection. JavaScript obfuscation can transform a known malicious sample into an undetectable one. On top of this, existing bundlers in the ecosystem (e.g., Webpack) enable the mixing of benign and malicious scripts, making static analysis harder. The other side of obfuscation is evasions: in addition to making code comprehension (via human or machine) harder, malicious actors have deployed time bombs, offloading parts of the malicious script to be read from the DOM or via a network request, avoiding detection.~\cite{fv8-sec24}

\subsection{Hierarchical clustering}
This paper utilizes hierarchical clustering, an unsupervised learning technique for segmenting data into nested structures (clusters) to identify pages that share phishing kits from their client-side behaviors. Specifically, we use HDBSCAN, a hierarchical variant of the density-based DBSCAN clustering technique. Advantages of HDBSCAN include not requiring prior knowledge of the number of clusters and no $\epsilon$ hyper-parameter tuning required out of the box\footnote{HDBSCAN picks the most stable clusters using the excess mass algorithm}. HDBSCAN has been used by prior work for categorizing malware families~\cite{fv8-sec24, Kizzle}.

When we have ground truth, we use the V-measure and Fowlkes-Mallows index (FMI) to validate our clusters. Both are scores of 0 or 1 that indicate how well the clusters map to ground-truth classes. V-measure (Validity measure) is the harmonic mean between completeness (all members of the same class are clustered together) and homogeneity (clusters only contain members of a single class)~\cite{Rosenberg2007VMeasureAC}. FMI, on the other hand, is the geometric mean between precision and recall, providing a close analog for an F1 score in supervised learning.
When we do not have ground truth for the pages, we use the silhouette score of all the clusters. The silhouette score measures how well separated the clusters are, with values ranging from -1 to 1. While it is usually used to fine-tune hyperparameters, we primarily use it to measure how well-formed the structures we extract from HDBSCAN are for all phishing pages. Scores under zero signal overlapping structure, while scores above 0.5 and 0.7 indicate medium or firm separation, respectively. 

%% file: tex/methods.tex
\section{Methodology}
\label{sec:methods}
\input{figures/infra}
This paper provides a methodology for clustering based on dynamic features, evaluates how closely the clusters resemble underlying phishing kits, and describes the prevalence of different adversarial techniques in the ecosystem. The building blocks of our experimental design are browser API execution traces from phishing pages and a ground-truth dataset of pages where we know the underlying phishing kit. In the following section, we describe the experimental setup for collecting this data, the steps we took to aggregate and enrich the execution traces, how we annotate the clusters based on the techniques used, and, finally, the steps we took to cluster the data and evaluate them as analogs for phishing kits. Figure~\ref{fig:infra} shows an overview of our pipeline.
 
\subsection{Collecting Traces and Kits}

Our crawling infrastructure ingests phishing URLs, when applicable, labeled with the entity being targeted from phishing feeds, and outputs execution traces for each page, along with the potential kit used for that page.

\myparagraph{URL feeds} We gathered phishing urls by monitoring the following phishing feeds for 17 months, from August, 2023 to January, 2025.: OpenPhish~\cite{openphish}, PhishTank~\cite{phishtank}, SMS Gateways~\cite{nahapetyan2024sms}, URLScan~\cite{urlscan}, PhishDB~\cite{phishingDB}, and APWG~\cite{apwg}, based on the availability of the feed and the level of access we had at the time.
Every two hours, we checked these feeds for new URLs (limited to the last 48 hours) and submitted them to two different crawlers: VisibleV8 and \kp{}. When available from APWG and OpenPhish, we also store the user-reported entity for the brand being impersonated. 

\myparagraph{VisibleV8} To get execution traces for every script loaded when visiting the page, we used an automated Chromium-based crawler with VisibleV8 patches applied~\cite{vv8-imc19}. The VisibleV8 patches modify Chromium to log all JavaScript APIs executed for every script loaded on the page. We automate the browser to visit the page and take screenshots using puppeteer~\cite{pptr}, an npm package by Google for browser automation. The patched Chromium crawler uses puppeteer-stealth, a set of configurations to help mask the headless Chrome and Puppeteer itself from detection tools~\cite{puppeteerstealth}. We initiated the crawls from a network designated for research purposes, for which the ISP would register as `educational' for any IP intelligence API. We used Catapult~\cite{catapult}, a man-in-the-middle proxy, to capture the entire HTTP archive for replayability. The crawler stays on the page for 45 seconds before taking a screenshot, allowing scripts to load and start executing, consistent with prior work~\cite{vv8-imc19,englehardt2016census}.

\myparagraph{\kp{}} While not guaranteed, some malicious actors leave the zip files of the kits used in a discoverable folder on the same server that hosts the website (for example, the Apache document root). \kp{}~\cite{kitphishr} is a Go-based URL fuzzer that attempts to identify any leftover zip files on the server. Prior work~\cite{liu_inferring_2022,oest_inside_2018} establishes \kp{} as a method for collecting and analyzing phishing kits. If successful, it will download the zip file and make a note of the domain from which \kp{} acquired it.

\myparagraph{Brand's original page} OpenPhish and APWG report the brand that a phishing page targets. We selected 5 of the top 52 brands targeted based on popularity by page number and on brands that represented seasonality-targeted sectors (e.g., IRS or banks). We collected VisibleV8 logs for their home pages and, when applicable, their login pages.

\subsection{De-duplication}

Once we have collected browser API traces and potential phishing kits, we post-process the traces into a set of APIs executed in a first-party context per page and de-duplicate the phishing kits.

\myparagraph{Trace postprocessing} We build our queries on top of the VisibleV8 Mfeatures postprocessor. These programs take the raw logs generated by the patched Chromium browser and convert them into an organized database, identifying duplicate scripts via SHA3 hash, clearly marking the origin of each script that loaded, and isolating which JavaScript API calls are browser API calls defined in the WebIDL file\footnote{\url{https://developer.mozilla.org/en-US/docs/Glossary/WebIDL}} generated while building the patched Chromium. For our analysis, we extracted tuples of the original page's URL, the script's URL, and an unordered set of APIs executed by this script.

To avoid introducing artifacts into our clusters from cloaked pages and third-party scripts, e.g., Google Analytics, known to be present on phishing pages~\cite{phishingvLegit}, we isolate API sets executed by first-party scripts (henceforth called first-party API sets). We set the root domain to the domain submitted to the feeds, and the origin to the domain from which the script is loaded. We consider a script first-party only if it is hosted on the same domain we acquired from our feeds (root domain). The only exception is when we identify which pages embed a CloudFlare Turnstile script. As some of the Turnstile scripts are hosted on `CloudFlare.com.'

\myparagraph{Identifying kit families} We crawl the URLs with \kp{} to establish a ground truth dataset with URLs originating from the same kit\footnote{We discard all domains that yielded two or more zip files}. For zip files extracted via \kp{} that have password-based encryption enabled, we use the SHA256 hash of the zip files to identify which domains yielded the same kit. For the remaining zip files, we further deduplicate them into kit families, treating them as sets of SHA256 hashes of source files~\footnote {Source files identified via a Python Magika module~\cite{fratantonio25:magika}}. If the Jaccard index-based similarity, where $JI(A, B) = |A \cap B| / |A \cup B|$ of these sets is equal to or greater than 80\% (selected to be conservative in aggregation), we consider the two kits to be from the same family and thus group all domains from both kits into the same group.
To ensure that common configuration changes do not separate two kit families, we remove code comments, emails, phone numbers, IP addresses, and URLs from the source files using regular expressions, replacing them with static identifiers. If we identify the source file as a blocklist because 90\% of it consists of IP addresses, we remove it entirely from consideration, since the blocklist can change between deployments. 

\myparagraph{Adjusting for CloudFlare} Many anomalies (multi-layer evals, non-deterministic behavior, dynamically generated scripts) originate from CloudFlare scripts on the same domain as the phishing pages. Since CloudFlare scripts load from a URL containing 'cdn-cgi' in the path, we do not consider scripts loaded from that endpoint in our analysis, except when we identify Turnstile usage in phishing pages.

\subsection{Identifying kits}

We hypothesize that similarity in the execution of browser APIs indicates that the pages originate from the same phishing kit. To test this, we ingest API sets from pages we identified as containing phishing kits and output a potential clustering of those pages, checking how well the clustering maps back to the ground truth.

To establish this similarity, we use the Jaccard index on the first-party API sets with Hierarchical Density-Based Spatial Clustering of Applications with Noise~\cite{hdbscan} (HDBSCAN), with a minimum cluster size of 2, to cluster the pages with the Jaccard distance as our distance kernel. Since HDBSCAN requires only a minimum cluster size and no prior knowledge of the number of clusters, we use the ground truth labels from \kp{} to evaluate the clustering. When ground truth is available, we evaluate the clustering using the \emph{Fowlkes-Mallows Index}~\cite{fowlkes1983method} and \emph{V-measure}~\cite{Rosenberg2007VMeasureAC}. We use sklearn's measure module to calculate all cluster evaluation metrics~\cite{scikit-learn}, and separate all the noise elements into singleton clusters for evaluation, as removing all noise samples would give it an unreasonable advantage; however, keeping the elements in the same noise cluster would artificially reduce the homogeneity score. For the unlabeled set of pages for which we do not have kit labels, we use silhouette score\footnote{While silhouette score is biased against non-convex clusters, based on our results, we do not see it necessary to switch to a density-specific cluster metric}, a metric for how well packed and separated the clusters are, to evaluate the clustering when we do not have ground truth.

To process \totalPagesClusterable{} pages would require a distance matrix of distances (64-bit float) over a 1.5Tb in size. To resolve this restriction, we first divide our data into 4-week rolling windows, advancing by 2 weeks, and cluster pages within each window. We refer to these clusters as `local clusters'. However, these local clusters cannot represent phenomena like the re-emergence of kits if it occurs after 2 weeks. 
To capture re-emergence, we isolate a representative set of APIs for the local cluster, as the intersection between all the cluster members' API sets. Once we create a representative set of APIs for each local cluster, we use DBSCAN with a conservative $\epsilon = \dbscanEpsi$ to merge local clusters whose representative sets are highly similar. This $\epsilon$ was selected because it yielded the best FMI across the pages used in the ground-truth evaluation for the large mega-clusters.

\subsection{Data enrichment}
\input{figures/behavior_categories}

In addition to all the data gathered, our analysis references a manually crafted Browser API-to-phishing-technique mapping and a characterization of cluster lifetimes and deployment diversity.

\myparagraph{Technique to Browser API mapping} JavaScript code can engage in data harvesting (exfiltrating information dynamically and not through form submission), evasion (conditional dynamic behavior aimed at hiding functionality or contents), obfuscation (unconditional behavior meant to hinder static analysis), and mimicking (dynamic behavior to make the page more believable, for example, false loading pages, stage by stage data extraction). Based on prior work by Su~\etal{}~\cite{jsufp} and Zhang~\etal{}~\cite{zhangCrawlPhishLargescaleAnalysis2021} and manually identifying APIs from the Mozilla Developers Network (MDN) documentation, we present a table mapping standard phishing techniques to browser APIs in Table~\ref{tab:behaviorcategories}. We leverage the presence of these APIs in the execution traces as a signal of the technique being present in the page. If the page falls within a specific cluster, we mark the entire cluster as using that technique. We use this to combat the phishing pages that engage in non-deterministic behavior. For CloudFlare Turnstile embedding, we use a non-browser API as our detection metric, as CloudFlare's native turnstile script will read the value of \textit{Window.Turnstiles}.
For Client-side IP checks, we manually looked through every \textit{Window.fetch} and \textit{XMLHttpRequest.open} argument URL given that argument was present in more than 50 phishing pages, and identified 14 that were IP reputation APIs. 

\myparagraph{Cluster lifetime and deployment diversity} Throughout this work, we refer to cluster lifetime as the time range between when we observe the first page belonging to the cluster on a phishing feed and when we observe the last page belonging to the cluster. 
We measure deployment diversity of the phishing pages by looking at the effective top-level domain plus one (eTLD+1) for the URLs. Using the eTLD+1 instead of the entire hostname ensures that pages deployed on 'pages.dev` or 'blogger.com` are considered a single deployment form.

%% file: figures/infra.tex
\begin{figure*}[t]
    \includegraphics[width=\textwidth]{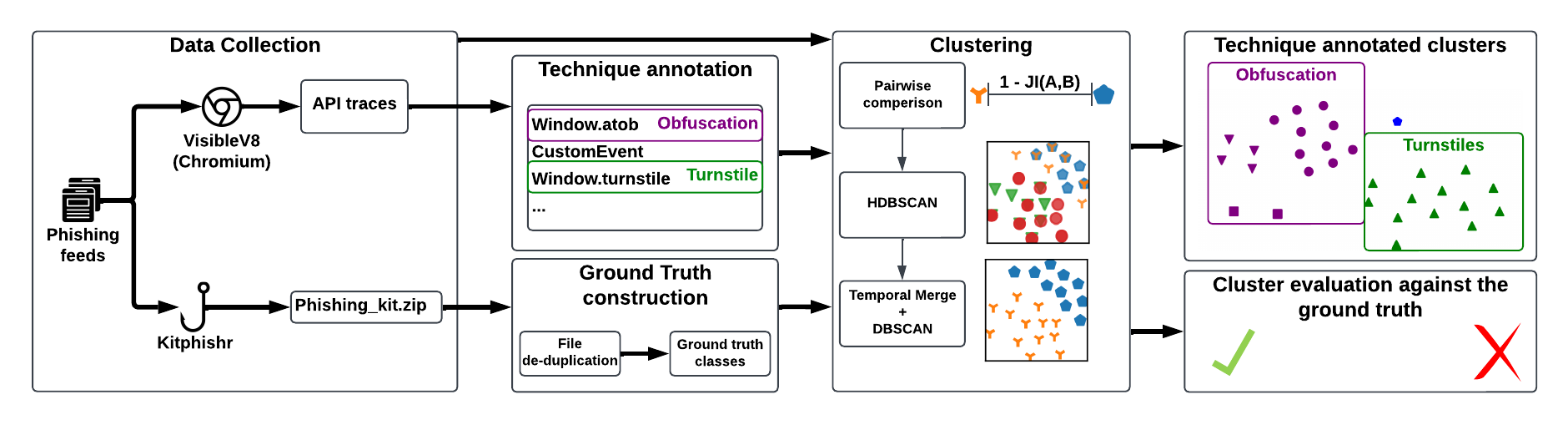}
    \caption{Crawling and clustering infrastructure}
    \label{fig:infra}
\end{figure*}

%% file: figures/behavior_categories.tex
\begin{table*}[t]
    
    \caption{Manual mapping of phishing techniques to browser APIs}
    \resizebox{\textwidth}{!}{%
    \begin{tabular}{l|l|p{10cm}}
        \textbf{Technique} & \textbf{Category} & \textbf{Identifying markers} \\
        \hline
        Fingerprinting extraction & Credential Harvesting & 10 Fingerprinting API calls, and an exfiltration-related API call from ~\cite{jsufp} \\
        \hline
        Client-side IP check & Evasion & 
        \texttt{Window.fetch} \newline
        \texttt{XMLHttpRequest.open} \\
        \hline
        Timing bot detection & Evasion & 
        \texttt{Performance.now + Timeout} \\
        \hline
        Encryption & Obfuscation & 
        \texttt{SubtleCrypto.decrypt} \\
        \hline
        Encoding & Obfuscation & 
        \texttt{TextDecoder.decode} \newline
        \texttt{window.atob} \\
        \hline
        Dynamic script Evaluation & Obfuscation & 
        \texttt{eval} \\
        \hline
        Basic fingerprinting & Evasion & 
        \texttt{HTMLDocument.cookie} \newline
        \texttt{HTMLDocument.referrer} \newline
        \texttt{Navigator.userAgent} \\
        \hline
        Dynamic script creation & Evasion & 
        \texttt{HTMLScriptElement.text} \newline
        \texttt{HTMLScriptElement.innerHTML} \\
        \hline
        CloudFlare Turnstile & Evasion & 
        \texttt{Window.turnstile} \\
        \hline
        Pop-ups & Evasion & 
        \texttt{Navigator.requestMIDIAccess} \newline
        \texttt{Clipboard.readText} \newline
        \texttt{Geolocation.getCurrentPosition} \newline
        \texttt{MediaDevices.getDisplayMedia} \newline
        \texttt{HID.requestDevice} \newline
        \texttt{Window.\{confirm|alert|prompt\}} \newline
        \texttt{Accelerometer|Gyroscope} \newline
        \texttt{Window.showModalDialog} \newline
        \texttt{MediaDevices.getUserMedia} \newline
        \texttt{SyncManager.register} \newline
        \texttt{Clipboard.read} \newline
        \texttt{Serial.requestPort} \newline
        \texttt{USB.requestDevice} \newline
        \texttt{Window.queryLocalFonts} \newline
        \texttt{Notification.requestPermission} \\
        \end{tabular}
        
    }
\label{tab:behaviorcategories}
\end{table*}

%% file: tex/data.tex
\section{Data characterization} 
\label{sec:data}
This paper utilizes two distinct datasets: a labeled dataset of phishing pages deployed in the wild with their corresponding phishing kit, and an unlabeled dataset of phishing pages. 

We support the use of our data and tools by other researchers for any follow-up study of the phishing ecosystem or reproducibility work.
However, due to ethical concerns highlighted in the appendix, the data will be available for researchers \emph{upon request}. Besides containing URLs shared through data agreements with the feed operators, collected screenshots, HAR archives, and VisibleV8 logs amounting to 6.3TB of data, which could contain identifiable information about the submitters in packed binary archives (catapult's HAR files), and create a heavy load on any data-sharing platform used.

\subsection{Labeled dataset}
\myparagraph{Overview} The ground truth data for this paper is a mapping of \totalNumberOfKitFamilies{} kit families to \totalKitPages{} phishing pages. We collected \totalZipFiles{} archive files by running all \totalPages{} through \kp. We discarded two encrypted zip files as we could not accurately link them to any of the other kits present. We further filtered the archive files to ensure they contained at least one code file, as Magika~\cite{fratantonio25:magika} identified. This reduces our archive count to \totalNumberOfKits{}. Finally, we group these kits into  1,328 families out of which \totalNumberOfKitFamilies{} families executed least 4 browser APIs. However, we examine other thresholds for selection in Section~\ref{sec:results}. 

The most-deployed kit was `2e94aff28a2c', a kit targeting Wells Fargo (1,073 URLs), which made use of 3 separate server-side blocklists (.htaccess file and 2 PHP modules with regex rules for user-agent and IP addresses) and called 52 distinct browser APIs. Another kit of interest was seen from 256 distinct URLs was ``fce61e98018d'', a USPS phishing kit which executed on average 170 distinct APIs, with server-side and client-side IP checks, advanced fingerprinting API calls, obfuscation, and \js{} generated DOM (Vite). 

In line with prior work~\cite{oest_inside_2018, divakaranPhishingDetectionLeveraging2022}, 89\% of these phishing kits were written in PHP, and we found 11 kits contained Python code. During our manual analysis of these kits, we concurred with prior work that many of them reused each other's code, especially regarding server-side IP blocklists.

Supporting our hypothesis, phishing pages from different kit families have vastly different browser API usage. Pages from different kits have an average API similarity of 15.9\% ($\sigma=0.2$). On the other hand, on average, the pages from the same kit family have an API similarity of 98.6\% ($\sigma=0.1$). After manual examination, we found that in cases where the kit yields pages with dissimilar API usage, the culprit is different paths within the domain being visited. 

\myparagraph{Kit-binding} By viewing the binding between the kit and a page as a distribution of what percentage of filenames loaded from the root domain appear in the kit's zipfile, we find that for 2,889 pages this falls between the 90-100\% match against the kit collected. We note that a 0\% match should not be treated as an indication of failure to bind the kit to the visited page, but rather an inconclusive result. We show the distribution of this similarity in Figure~\ref{fig:kitbindbing}. The most widely deployed non-binding-kit was a single index.php file.

\subsection{Unlabeled Phishing pages}
\input{figures/median_distance_of_clusters.tex}
In total, we collected browser traces from \totalPages{} URLs. These included false Facebook suspension, fake USPS mail delivery notifications, tech support scams, AT\&T login pages, and Bitcoin Wallet logins. All the pages collected have been labeled as phishing, as they pose as a trustworthy entity to gain credentials or network access from their victims, despite their varying tactics and credentials collected~\cite{CISA}. 
388,536 pages did not execute any APIs in the first-party context, and only \totalPagesClusterable{} executed at least 8 APIs in a first-party context to qualify for clustering, with the remaining pages not executing enough distinct browser APIs to be considered for clustering. 

We first processed the pages into 36 clusterings (groups of the local clusters). Figure~\ref{fig:localClusters} shows the distribution of the silhouette score of these local clusters. With an average score of 0.83, these clusters are well-formed. Once clustered together, the average cluster in our dataset contained 47 pages ($\sigma=405$), lasted 39 days ($\sigma=74$), and had 66 APIs ($\sigma=57$) in common across all member pages. We remove \badClusters{} local-clusters from the merge algorithm, as they formed ``malformed clusters'' with fewer than 4-APIs in common between all pages.

After manually inspecting a sample of clusters, we observed that these clusters have unique pages across cloud service providers and languages. For example, Cluster-98b0e6bb (shown in Figure~\ref{fig:ms_defender}) comprises 487 pages across five unique eTLD+1. It contains pages that engage in voice-based phishing attacks in Japanese and English, with varying phone numbers and error messages. With over 57 APIs in common, it is clear that these pages' use of keyboard-interception APIs, Audio APIs, and Network APIs (to ipwho.is) for IP intelligence caused them to cluster together.

%% file: figures/median_distance_of_clusters.tex
\begin{figure}[ht]
  \includegraphics[width=\columnwidth]{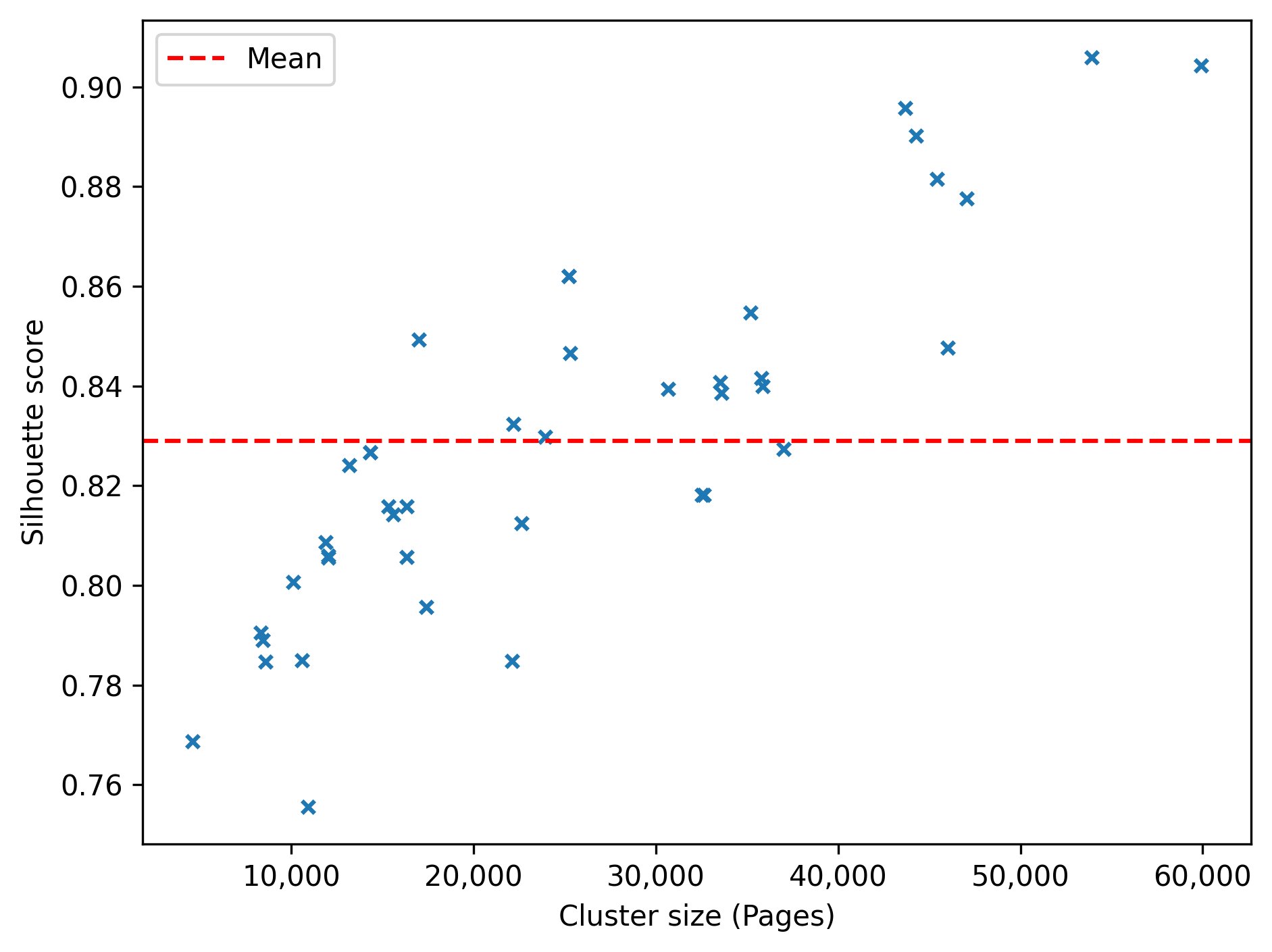}
  \caption{Distribution of the silhouette score of local clusters compared to their size in number of pages.}
  \label{fig:localClusters}
\end{figure}

%% file: tex/results.tex
 \section{Results} 
\label{sec:results}

\subsection{Kit identification}
\input{figures/vt_vs.tex}
\paperFinding{In a small population size (\totalKitPages{} pages) that execute at least four distinct browser APIs, pages can be related to one another based on the underlying phishing kit} We find that clustering all pages that execute at least two browser APIs yield an FMI of 0.92. We note that as the sophistication of the page increases, so does the clustering accuracy. Figure~\ref{fig:vs_gt} shows the V-Measure and FMI for our clusters as we increase the requirement for distinct APIs in the execution trace. We ultimately chose four browser APIs as the requirement for further experimentation on the ground truth data, as they provided a good tradeoff between V-measure and the number of pages used. For the unlabeled pages, we used 8 APIs, as they produced no malformed clusters in our ground truth data (i.e., clusters with no API sets in common).
Clustering pages from \totalKitPages{} pages across \totalNumberOfKitFamilies{} kits yields \totalKitClusters{} clusters. Evaluating these clusters against the ground truth labels for each page, we find that our clusters have an FMI-based accuracy of \gtFMI{} and a V-measure of \gtVS{} (with completeness and homogeneity scores of \gtcompleteness{} and \gthomogeneity{} respectively).

The kit `2e94aff28a2c' targeting Wells Fargo accounts for 26\% of the ground-truth dataset. We evaluate clustering on a rebalanced dataset, where we sample 104 pages (the same number as the second-most-popular kit) instead of using all 1,073. This reduces our FMI score to 0.73 (V-Measure of 0.91), maintaining higher accuracy than prior work. Manual examination of this accuracy drop reveals that, in many cases, this result stems from different paths within the same domain not being clustered together, as would be the case with multi-stage phishing pages, where different stages were submitted to our feeds. 
We discuss avenues to improve this accuracy in Section~\ref{sec:limitations}.

\input{figures/static}
\paperFinding{Browser API sets better separate phishing kits than script hashes, even when we include scripts dynamically extracted from eval statements} We use SHA256 hashes of executed scripts as features in HDBSCAN to compare against our approach. Table~\ref{tab:static} shows that browser API sets maintain a higher accuracy than SHA256 hashes of scripts, even for scripts that are extracted out of eval statements, and would require dynamic or static analysis to acquire.

\paperFinding{DOM APIs and property reads are a valuable signal in kit differentiation} We find that removing DOM-related APIs or property reads out of consideration drastically reduces the number of pages we can consider for ground truth evaluation without significantly increasing our overall accuracy. The key insight here is that these APIs signal particular choices the kit author made about the UI library they used: if they chose to hide the DOM as an evasion, or not draw it to begin with, do they refer to DOM elements by IDs, classes, or tags. Clustering evaluation metrics for the ground truth dataset with DOM, SVG, and CSS APIs removed results in an FMI of 0.88 and a V-measure of 0.86. After removing all property reads, we observed FMI and V-measure values of 0.98 and 0.91, respectively. In both cases, we can cluster fewer pages and thus identify fewer kits, but do not see an improvement in clustering accuracy.

\subsection{Clusters in the wild}
\paperFinding{ Clusters primary are formed from pages targeting the same brands, with 69\% of clusters containing URLs only marked by a single target brand by our threat intel sources\footnote{We did not include clusters in this count that had no brand-labeled URLs in them.}} This phenomenon is observed in the ground truth dataset, where URLs for 90\% of the kits targeted a single brand. 

While we observe multi-branded clusters, such as the one shown in Figure~\ref{fig:example2}, the majority of kits are single-branded, consistent with prior work examining collected kits in the wild~\cite {leePhishingScriptsScriptLevel2024}. Together, this provides strong circumstantial evidence that deployed kits are increasingly becoming brand-specific. On the other hand, 1,047 clusters (18\%) had two brand labels. However, the most popular combinations were "Meta/Facebook", "Facebook/Instagram", "National Police Agency JAPAN/Microsoft", and "Facebook/WhatsApp", keeping the parent organization of the target the same in the majority of cases. Manual examination of clusters with the "National Police Agency JAPAN/Microsoft" brand label revealed that shopping pages in Japanese were mismarked with that label in our data feeds.

\paperFinding{UI interactivity and fingerprinting are a near-universal behavior across clusters} Multi-stage phishing pages are very well documented in prior work, and we find that most clusters (92\%) register a click event listener. Though this could be as simple as submitting credentials via JavaScript, it highlights the need for researchers to augment their crawlers in the future to extract better, more complete execution traces from websites. We split fingerprinting into two categories, basic and advanced. Basic fingerprinting, which follows the list of APIs identified by Zhang~\etal{} in ~\cite{zhangCrawlPhishLargescaleAnalysis2021} was present in 80\% of the clusters (289,954 pages), and advanced fingerprinting (measured by at least 5 APIs identified by Su~\etal{} in ~\cite{jsufp}) showed up in 70\% of the clusters.
\input{figures/obfuscation_tactics.tex}

\paperFinding{Fingerprint exfiltration, dynamic script creation, obfuscation, and bot detection are uncommon across clusters} While fingerprinting is near universal, we find that a smaller fraction of the clusters employ obfuscation, fingerprint exfiltration, and timing for bot detection. Prior work has shown interest in these behaviors~\cite{sanchez-rolaRodsLaserBeams2023,fv8-sec24,linPhishSheepClothing2022,zhangCrawlPhishLargescaleAnalysis2021}, meaning kits that forgo this may be rudimentary either by negligence or design, to avoid detection. Dropping anti-bot detection features has been observed before, with an Office-365 phishing kit (Tycoon2FA), opting to remove CloudFlare Turnstile integration, as it was being used as a feature for detection~\cite{Tycoon2FANewEvasion}.

Another common tactic for bot detection is timing-based checks: by calling \textit{Performance.now} and \textit{Window.setTimeout} statements to measure the time differential between setting the timeout and its triggering. We find that 22\% of clusters call \textit{Performance.now} in conjunction with \textit{setTimeout}.
\input{figures/example_embedding.tex}
On the other hand, 21\% of clusters (1,991) employ some form of obfuscation, and 1,092 of the clusters (16,859 pages) dynamically generate an HTML script element. Table~\ref{tab:obfuscaiton} lists all obfuscation techniques, with eval and Base64 encoding as the most popular methods. Despite the best recommendations to web developers to avoid using `eval'~\cite{EvalJavaScriptMDN2025}, JavaScript's eval function remains a favorite for obfuscation and evasions~\cite{fv8-sec24}. We observe script executed inside an `eval()', evaluating yet another script; we measure this phenomenon as a level in \emph{eval-depth}. We find that 48 clusters have pages that go to eval-depth 3. However, this seems to be a side-effect of embedding the phishing pages (mainly ones targeting Facebook) in Blogger.com pages.

\input{figures/ipinfo.tex}
\paperFinding{While rare, client-side IP reputation checks are present across multiple clusters} While only present in 412 clusters (19,778 pages), we identify 14 unique IP reputation APIs used by phishing pages as soon as the page loads. We present a complete breakdown in Table~\-\ref{tab:ip_rep}. While not the most popular, \textit{api.ipregistry.co} presents an interesting case study, as it enables the identification of educational networks. Manual examination of pages from these clusters reveals snippets similar to Listing~\ref{fig:ip_example}, which conditionally chooses to redirect away from cloud hosting providers, content delivery networks, and educational networks, like the one through which we perform the crawls. However, because they employ other browser APIs, we can still cluster the pages based on the landing page's initial logic before the cloaking behavior.

\paperFinding{Pop-Up APIs are declining in usage} We see only 88 clusters (1,331 pages) call out to pop-up requesting APIs. Geolocation.getCurrentPosition (44 clusters) was the most popular among these. While it requires a pop-up to interact with, this API can also be crucial for cloaking, as any VPN or proxy cannot mask the results. We observe a smaller fraction of the ecosystem (13 clusters, 151 pages) than \cite{zhangCrawlPhishLargescaleAnalysis2021} employs this cloaking technique, especially when it comes to triggering a notification pop-up to verify user interaction. This could be a result of Firefox, citing low engagement with the notifications, starting to require user interaction to trigger the popup~\cite{mozillaRestrictingNotificationPermission2019} at the end of November 2019, when Crawlphish's data collection ended. Chrome has since discussed modifying the notification API to make the request less disruptive to the user experience~\cite{IntroducingQuieterPermission}. The lack of pop-up requests could also be explained by our \emph{finding 7} regarding the usage of IP intelligence APIs or by the overwhelming amount of clusters (91\%) registering at least one HTMLElement event handler, which would be classified as a \textit{Click-through} by Crawlphish's taxonomy. We report a full breakdown of APIs related to the Crawlphish categorization of client-side cloaking in Table~\ref{tab:crawlphish}.

\input{figures/crawlphish_all_categories.tex}

\paperFinding{Mouse Detection API calls and CloudFlare Turnstile embedding are specific to a small group of clusters}
While supported by most modern browsers, we observe an infrequent use of Mouse Detection APIs for bot detection. Only 28 clusters use mouse-detection-related APIs. Two of these clusters are from an open-source phishing kit, leveraging BotGuard, and are from a public GitHub repository, which was last updated in 2017\footnote{\url{https://github.com/ashanahw/Gmail_Phishing}} shown in Figure~\ref{fig:example2}. We see these clusters deployed across 17 unique domains, from 2023-10-07 to 2024-07-19. 

Although mentioned in prominent campaigns~\cite{SneakyPhishJust2025}, only \emph{9 clusters (184 pages)} embed a CloudFlare Turnstile check; some domains are not hosted on CloudFlare. Embedding a CloudFlare Turnstile check allows the phishing kit authors to offload bot detection to a well-established ecosystem. Since 2024, high-value phishing kits have been identified with this behavior~\cite{Tycoon}. However, subsequent analysis of the same threat actor identified a shift from turnstiles to HTML-Canvas drawn captchas. 
\input{figures/ip_apis.tex}

\paperFinding{Clusters that utilize both experimental browser APIs, and extremely deprecated APIs} We find 320 clusters that use \textit{NavigatorUAData.getHighEntropyValues} for fingerprinting at the time of data collection, an experimental API not fully supported by Firefox and Safari. While using experimental APIs for fingerprinting is common, we observe use cases for mimicry and social engineering. Of the 136 clusters that used \textit{Keyboard.lock}, the cluster shown in Figure~\ref{fig:ms_defender} in the appendix left the user unable to press keys. 

We identified 9 clusters across 4,315 pages that used \textit{Scheduling.isInputPending}; however, upon closer inspection, these were not pages using a novel kit, but pages that used Google Presentation to construct their landing page. On the other hand, 24\% of the clusters spanning (47,189) pages use a deprecated API.

Figure~\ref{fig:AllTechniques} shows the confusion matrix between the techniques we enumerated and the size of the clusters. We see no noticeable difference in techniques regarding cluster size, except for a higher percentage of large clusters using client-side IP checks. We also observe that pages that use experimental APIs tend to include a deprecated API call, which aligns with their use for browser fingerprinting rather than novel cloaking logic. 

\paperFinding{Phishing pages vary wildly from the brand's login page that they are mimicking}
We collect browser API traces from the login pages of Facebook, USPS, Meta, Microsoft, and the IRS and compare them to their phishing counterparts. The average similarity of the APIs executed by the phishing page and the original page is 11\%, indicating that browser APIs do not relate to the target page. We found no pages where the original page's API set was a subset of the phishing page's API set, and, in 5\% of the cases, the phishing page executed at least half of the APIs from the original page. We report per-brand findings in Table~\ref{tab:phishing-metrics} and note that the least similar brand was USPS, which could be the result of USPS phishing pages being multi-stage pages requiring user interaction and targeting credit card information~\cite{centerUSPSPhishingScam}.

%% file: figures/vt_vs.tex
\begin{figure*}[t]
  \includegraphics[width=\textwidth,trim=0 0 0 0]{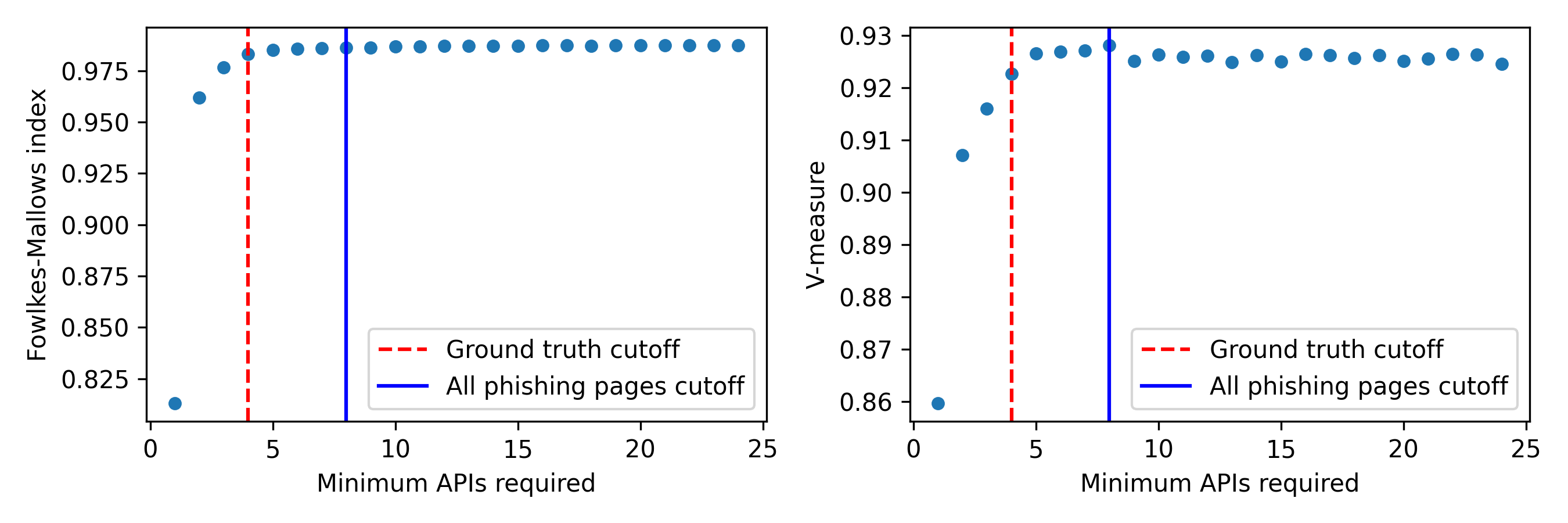}
  \caption{Validity measures for clusters vs. minimum distinct APIs required for clustering}
  \label{fig:vs_gt}
\end{figure*}

%% file: figures/static.tex
\begin{table}[t]
\centering
\caption{Comparison of the evaluation metric when script hashes are used instead of APIs executed}
\label{tab:static}
\resizebox{0.8\columnwidth}{!}{%
\begin{tabular}{lcc}
\toprule
\textbf{Method} & \textbf{FMI} & \textbf{V-measure} \\
\midrule

dynamic & 0.98 & 0.92 \\
Scripts (No eval) & 0.82 & 0.82 \\
Scripts (No eval, 1st party) & 0.75 & 0.79 \\
Scripts (1st party) & 0.71 & 0.77 \\
All script & 0.85 & 0.84 \\

\bottomrule
\end{tabular}
}
\end{table}

%% file: figures/obfuscation_tactics.tex
\begin{table}[t]
\centering
\caption{Breakdown of the obfuscation techniques observed in our dataset}
\label{tab:obfuscaiton}
\resizebox{0.8\columnwidth}{!}{%
\begin{tabular}{lrr}
\toprule
\textbf{Obfuscation techniques} & \textbf{Pages} & \textbf{Clusters} \\
\midrule

\texttt{Window.atob} & 61,125 & 1,455 \\
\texttt{eval} & 14,561 & 982 \\
\texttt{TextDecoder.decode} & 11,113 & 534 \\
\texttt{SubtleCrypto.decrypt} & 1,185 & 36 \\

\bottomrule
\end{tabular}
}
\end{table}

%% file: figures/example_embedding.tex
\definecolor{customblue}{HTML}{1f77b4}
\definecolor{customorange}{HTML}{ff7f0e}

\begin{figure}[t]
    \includegraphics[width=\columnwidth,trim=0 0 0 0]{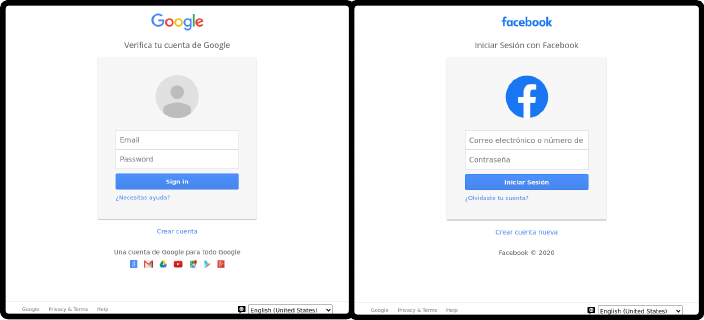}
    \caption{Screenshots of the two variants of a page found in the cluster associated with a phishing kit last updated in 2017 are still available on GitHub.}
    \label{fig:example2}
\end{figure}

%% file: figures/ipinfo.tex
\begin{listing}[t]
    \centering
        \includegraphics[
        width=\linewidth,
        trim=0 0 0 0,  %
    ]{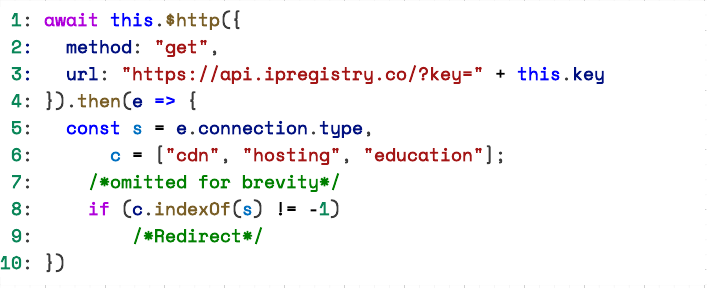}%
\normalsize
\caption{Example of IP-based cloaking by using a third-party reputation API. The following example specifically cloaks away from educational networks.}
\label{fig:ip_example}
\end{listing}

%% file: figures/crawlphish_all_categories.tex
\begin{table}[t]
\centering
\caption{Summary of all API calls that match to a category identified by Crawlphish}
\label{tab:crawlphish}
\setlength{\tabcolsep}{6pt}
\resizebox{\columnwidth}{!}{%
\begin{tabular}{lrr}
\toprule
\textbf{Category} & \textbf{Clusters} & \textbf{Pages} \\
\midrule

\textbf{User-Interaction} & & \\
\hspace{1em}Pop-up (Total) & 88 & 1,331 \\
\hspace{2em}\texttt{Accelerometer} & 5 & 63 \\
\hspace{2em}\texttt{Clipboard.readText} & 1 & 2 \\
\hspace{2em}\texttt{Geolocation.getCurrentPosition} & 44 & 477 \\
\hspace{2em}\texttt{Gyroscope} & 3 & 10 \\
\hspace{2em}\texttt{MediaDevices.getUserMedia} & 7 & 58 \\
\hspace{2em}\texttt{Notification.requestPermission} & 13 & 151 \\
\hspace{2em}\texttt{Window.alert} & 20 & 585 \\
\hspace{1em}Mouse & 28 & 60 \\
\hspace{1em}DomEvents & 8,540 & 364,794 \\

\midrule

\textbf{Fingerprint} & & \\
\hspace{1em}Total & 7,471 & 289,954 \\
\hspace{1em}\texttt{Navigator.userAgent} & 7,122 & 268,163 \\
\hspace{1em}\texttt{HTMLDocument.cookie} & 3,879 & 125,224 \\
\hspace{1em}\texttt{HTMLDocument.referer} & 4 & 10 \\

\midrule

\textbf{Bot Detection} & & \\
\hspace{1em}Timing & 2,528 & 78,202 \\

\bottomrule
\end{tabular}
}
\end{table}

%% file: figures/ip_apis.tex
\begin{table}[t]
\centering
\caption{List of all API endpoints that client-side code reaches out for IP intelligence}
\label{tab:ip_rep}
\resizebox{0.8\columnwidth}{!}{%
\begin{tabular}{lrr}
\toprule
\textbf{API url} & \textbf{clusters} & \textbf{pages} \\
\midrule

\texttt{api.ipify.org} & 143 & 7,415 \\
\texttt{freeipapi.com} & 31 & 6,441 \\
\texttt{api.ipregistry.co} & 8 & 3,994 \\
\texttt{api.db-ip.com} & 36 & 3,072 \\
\texttt{ipinfo.io} & 50 & 1,960 \\
\texttt{ipwho.is} & 38 & 794 \\
\texttt{get.geojs.io} & 11 & 753 \\
\texttt{ipapi.co} & 84 & 724 \\
\texttt{geolocation-db.com} & 14 & 492 \\
\texttt{geolocation.onetrust.com} & 38 & 368 \\
\texttt{api.ipapi.com} & 4 & 143 \\
\texttt{api.ipgeolocation.io} & 16 & 97 \\
\texttt{pro.ip-api.com} & 14 & 81 \\
\texttt{api.geoapify.com} & 3 & 63 \\

\bottomrule
\end{tabular}
}
\end{table}

%% file: tex/discussion.tex
\section{Discussion}
\label{sec:discussion}
\myparagraph{What makes a kit identifiable?}
The more sophisticated the phishing kit becomes, the easier it is to spot by the browser APIs it uses. Everything from exfiltrating a browser fingerprint to sophisticated evasions (e.g., canvas CAPTCHAs) makes mass deployments of the kit stand. There is also an economic incentive to sophistication, as novel use of browser APIs leads to better evasions from detectors~\cite{zhangCrawlPhishLargescaleAnalysis2021, oest_phishfarm_2019}, more valuable credentials harvested~\cite{sanchez-rolaRodsLaserBeams2023, linPhishSheepClothing2022}, or resilience to analysis~\cite{fv8-sec24,jsobf-imc20}. For example, the APIs \textit{Keyboard.lock}, \textit{HTMLDocument.onkeydown} for keyboard locking, \textit{Window.atob} for obfuscation, and a handful of fingerprint APIs and DOM APIs for dynamic content generation set the cluster shown in Figure~\ref{fig:ms_defender} apart from other pages. 

Browser APIs expose a suite of privileged behaviors to web pages, akin to system calls in an operating system. Our results indicate that, like system calls, these API calls serve as an excellent signal for aggregating phishing web pages. Browser API traces are resilient to obfuscation, as the APIs cannot be hidden from VisibleV8; if they are used by the page, they will appear in the trace. Even in the case of shared libraries, the only APIs that appear in the trace are those invoked by the page (either directly or via a library), meaning that the same library results in two different traces based on the functionality used.

\myparagraph{Advantages to behavioral aggregation}
Phishing feeds are a noisy data source for studying the ecosystem, from e-commerce pages to mass-spammed USPS and EZ-parking phishing pages. As client-side code for phishing pages becomes more complex, behavioral aggregation, akin to what prior work has done to identify exploit kits~\cite{Kizzle}, is necessary. The methodology in this paper is aimed at researchers and analysts. 

For research, identifying shared kits in a dataset of phishing pages helps control for easily obtainable or mass-deployed kits, enabling measurement of the prevalence of different techniques across kits rather than pages. \emph{At worst, we overestimate how popular different techniques are across kits, meaning we provide an upper bound for the less popular techniques.} For analysts, our methodology acts as a quick way to aggregate and share phishing kits-related threat intelligence between pages: things like server-side cloaking techniques, preferred exfiltration methods, ties to APTs, and data exfiltrated. While kit families can vary in which IPs they denylist and which user agents they allow, fingerprinting the underlying kit can help analysts determine whether a page employs these techniques in the first place.

As phishers leverage free website builders~\cite{saharoyPhishingFreeWaters2023} we note that our clustering aggregates phishing pages built using Weebly, Wix, and Google Sheets, respectively. While not created with dedicated crimeware (phishing kit), `living-off-the-land' pages that leverage free-website builders, as studied in ~\cite{saharoyPhishingFreeWaters2023}, can be trivially aggregated and isolated based on their dynamic similarity.

%% file: tex/related_work.tex
\section{Related Work} 
\label{sec:related}

\myparagraph{Phishing detection}
There is a wealth of research on phishing detection as the tug-of-war between adversaries and security professionals continues. Recently, Liu~\etal{} and Abdelnabi~\etal{} have deployed vision-based techniques to detect phishing pages ~\cite{liu_knowledge_2023,abdelnabiVisualPhishNetZeroDayPhishing2020,liu_inferring_2022}. ~\cite{liu_knowledge_2023} also presents over 6,000 phishing kits analyzed as part of the work. With adversarial attacks ensuring that the page looks different to crawlers and analysts, some have turned to extracting features from the URLs themselves, more recently via LLMs in ~\cite{Chiba2024DomainLynxLL,phishReplicant} and earlier via statistical models and machine learning in ~\cite{Shirazi2018Kn0wTD,Le2018URLNetLA,Rao2018DETECTIONOP,Verma2017WhatsIA}

\myparagraph{Studying and combating adversarial techniques} Divakaran~\etal{} in ~\cite{divakaranPhishingDetectionLeveraging2022} reaffirms the need to keep up with the latest adversarial techniques to build better detection systems for phishing. Prominent work in this area includes ~\cite{zhangCrawlPhishLargescaleAnalysis2021} by Zhang~\etal{}, which uncovered and categorized many novel client-side techniques by forcing the execution of phishing pages to trigger the cloaking behavior. Acharya~\etal{} in ~\cite{acharyaPhishPrintEvadingPhishing2021} uncovered that phishing pages can successfully evade blocklists by knowing how to identify their crawlers, and Oest~\etal{} in ~\cite{oest_phishfarm_2019} demonstrated that cloaking from non-mobile-based devices as a phishing page can ensure that the attacker's page goes unmarked by the blocklists for more than 48 hours.

Kondracki~\etal{} in ~\cite{kondrackiCatchingTransparentPhish2021} uncovered a massive blind spot of the phishing detection ecosystem that was man-in-the-middle phishing kits. Kits that would transparently forward the victim's connection to the target page, mimicking brand logos on pages like Outlook without any configuration. Fortunately, the authors addressed the blind spot by demonstrating that these proxies remain fingerprintable using TLS fingerprinting. ~\cite{tzschoppe_browser---middle_2023} proposed a similar attack using JavaScript and NoVNC to trick the user into signing in to their account via a VNC session in their browser.

As adversaries become more creative with their evasions and obfuscation techniques, some novel defenses have also opted to think outside the box. Zhang~\etal{} in ~\cite{zhangImSPARTACUSNo2022} proposed a phishing defense solution that leverages the high likelihood of a phishing page cloaking away from a crawler to the defender's advantage. They demonstrated that a web browser configured to look like a crawler triggers a cloaking response from phishing pages, ensuring that victims never see the page while maintaining compatibility with all of Alexa's top one million websites. Meanwhile,  PhishDecloaker~\cite{teoh_phishdecloaker_2024} utilized vision-based models to combat phishing pages. 

To better understand why phishing pages may choose to fingerprint, other than cloaking Lin~\etal{} in ~\cite{linPhishSheepClothing2022} showed that browser fingerprints could be successfully used to bypass multi-factor authentication, a system meant to be a last line of defense against stolen credentials, for 10 out of 16 websites that provide popular services.

\myparagraph{Phishing kits}
Much can be studied about the phishing ecosystem via phishing kits. Cova~\etal{} in ~\cite{covaThereNoFree} uncovered that most ``free'' phishing kits contain a backdoor, effectively serving as a way to offload the deployment of a campaign to a third-party while siphoning off their stolen credentials. 

Similar to our goals, PhiKitA~\cite{castanoPhiKitAPhishingKit2023} uses a dataset of phishing kits gathered through \kp{} and a collection of features extracted from the HTML DOM to classify websites into their matching kit. Their multi-class classifiers for identifying the kit achieved F1 scores of 39\%, 31\%, and 9\% across three algorithms. Merlo~\etal{} in ~\cite{merlo_phishing_2022} further expanded on our understanding of phishing kit lineages by looking at over 20,000 phishing kits and identifying, via token similarity, most of them as clones of one another or previously encountered kits.
Prominent work in extending our understanding of phishing attacks includes Han~\etal{} in ~\cite{hanPhishEyeLiveMonitoring2016}, where they monitor the deployment of phishing kits by adversaries that compromise vulnerable web servers by hosting a well-sandboxed honeypot. They collected 643 phishing kits and found that they can be installed and tested in minutes and remain undetected for weeks. Using these kits, they were also able to identify evasion techniques, such as path randomization per visit, which, at the time, was enough to bypass Google SafeBrowsing. 

In ~\cite{oest_inside_2018}, Oest~\etal{} manually analyzed phishing kits to establish the taxonomy for server-side cloaking, and in ~\cite{bijmansCatchingPhishersTheir2021}, Bijmans~\etal{}, after collecting phishing kits by watching TLS transparency logs to identify Dutch bank phishing domains, manually created a fingerprint from static features to analyze their prevalence in the wild. 

In 2024, Lee~\etal{} in ~\cite{leePhishingScriptsScriptLevel2024} provides a server-side script (PHP) level analysis of phishing kits, finding that dynamically generated URLs are still standard in the ecosystem and observing seasonality in the kits they were able to obtain. 

\myparagraph{Characterizing the phishing ecosystem}
Similar to our methods, Rola~\etal{} in ~\cite{sanchez-rolaRodsLaserBeams2023} deployed a modified Chromium browser to gather data and analyze the browser APIs of phishing websites, using a pre-selected API list focused on first- and third-party scripts used on phishing pages. They find that most of the most visited phishing pages (identified via browser telemetry data) deploy fingerprinting scripts, sometimes differing from those of the original brand they mimic. At the same time, they accessed this at the script level, reinforcing our finding that phishing pages vary widely from their original pages.

Oest~\etal{} in ~\cite{oest_sunrise_nodate} demonstrates the full lifecycle of a phishing campaign by employing the fact that phishing pages often copy assets from the target domain and refer the victim back to the original page afterward. By collaborating with a significant financial institution, they developed a framework to leverage this data and track a phishing page from its initial deployment to blocklists, flagging it as phishing. ~\cite{oest_sunrise_nodate} observed all techniques highlighted by prior work: cloaking, user-specific URL generation, man-in-the-middle proxies, and short-lived bursty attacks. Expanding our understanding of the victim experience on a phishing website, Subramani~\etal{} in ~\cite{subramani_phishinpatterns_2022} developed a crawler. 

\myparagraph{Dynamic analysis of webpages} Our work shares a methodology for dynamic analysis enabled by web measurement frameworks like OpenWPM~\cite{englehardt2016census} and VisibleV8~\cite{vv8-imc19}. Su~\etal{} used VisibleV8 traces and taint analysis in ~\cite{jsufp} to discover emerging fingerprinting techniques. Sarker~\etal{} used VisibleV8 to create an oracle for detecting obfuscation~\cite{jsobf-imc20}. Such an oracle was made possible by the observation that VisibleV8 marks the execution of an API at a given source line. At the same time, obfuscation techniques ensure that the API is not textually available there. Furthermore, Pantelaios~\etal{} used a combination of VisibleV8 and force execution modifications to the Chromium engine to identify and defeat JavaScript evasion techniques while also leveraging API traces and clustering to identify previously unlabeled malicious extensions~\cite{fv8-sec24}. Iqbal~\etal{} used OpenWPM to capture execution traces from Tranco's top 100K URLs, training a classifier on a mixture of dynamic and static features extracted from the JavaScript's AST and execution traces, respectively, to achieve a 99.8\% accuracy in identifying fingerprinting scripts online. Concurrently with this work, Favoretti \etal{} used browser API traces alongside two passes of DBSCAN to cluster pages from 409 kits, achieving a Normalized Mutual Information (NMI) of 0.85~\cite{sbseg}. Their approach used API sequences to approximate shared scripts across pages (1st DBSCAN) and then clustered pages that use the same scripts (2nd DBSCAN), requiring two distinct fine-tuned hyperparameters. Favoretti \etal{}'s underlying hypothesis is fundamentally different from ours: they use sequences of execution to identify duplicate scripts, whereas we use the presence of the APIs themselves (as proxies for data collection or evasion features). We also take a different approach to evaluation: they search the hyperparameter space for the optimal epsilon that maximizes their evaluation metric, while we do not fine-tuning on our evaluation metrics in our ground truth data, using FMI rather than NMI, which is prone to being overly optimistic in small cluster sizes as pointed out by Mahmoudi \etal{} in~\cite{mahmoudiProofBiasedBehavior2024}.

Our work differs from prior work in multiple ways. To date, we are the most successful at identifying phishing kits on a page; moreover, our approach is entirely automated, requiring no prior rule-based identification of kits. While we integrate many of the techniques annotated in prior work, we provide an up-to-date understanding of their distribution within their ecosystem and do so at the cluster level, which is more likely to control for multiple deployments of the same sophisticated kit. 

%% file: tex/limitations_future_work.tex
\section{Threats To Validity} 
\label{sec:limitations}
\myparagraph{Incorrect Ground truth mapping}
Prior work has used \kp{} with the intrinsic assumption that the kit collected is the deployed kit.
We note that a mismatch between the deployed and hosted kits would occur in cases of rare deployment errors by phishers or shared infrastructure. Because phishing infrastructure is ephemeral, averaging a lifetime of less than 2 days~\cite{oest_phishfarm_2019}, cases of shared infrastructure are expected to be rare; in most cases, this would result from two different entities compromising the same web host to serve their phishing kits.

\myparagraph{Unexplored Page states}
While we visit and loiter on phishing pages, we do not explore or interact with them. Addressing this limitation would only improve our cluster evaluation metrics. Upon manual inspection, this is a source of the clustering inaccuracy in the ground truth data, as some URLs point to different paths on the same kit (including specific stages of the phishing page or the root domain, which cloaks). With recent work on LLM-powered crawlers and ML-guided browser automation~\cite{subramani_phishinpatterns_2022}, we leave this to future work to collect a more complete list of browser APIs executed by a phishing page across all its stages.

\myparagraph{Third-party libraries} We use API sets rather than a sequence of APIs, executed, because we hypothesize that the kit developers' habits, along with the kits' required marketable traits, are encoded in the browser's API sets. This extends to the choice of libraries, their versions, and the functionality used. Libraries like jQuery and Lodash do not execute any APIs on their own when included. The APIs invoked in the trace result from kit-specific logic, using the libraries as an easy wrapper for features. Even hyper-specific libraries like AXIOS, a lightweight HTTP client for JavaScript, can differ in the APIs they use for network requests depending on the method calls and arguments provided.

%% file: tex/conclusion.tex
\section{Conclusion}
\label{sec:conclusion}
In this paper, we provided a methodology for researchers and analysts to differentiate phishing pages based on a common underlying kit automatically. We show an accuracy of \gtFMI{} on a dataset of collected pages and kits. By curating a mapping of techniques to browser APIs and \totalPagesClusterable{} pages, and identifying \totalClusters{} clusters, we explore which techniques are universal, widespread across kits, or kit-specific.

In doing so, we find that the majority of phishing pages between 2023 and 2025 that execute browser APIs make calls to fingerprinting APIs and rely on event handlers rather than direct mouse-detection API calls and pop-up-based cloaking. 
Our findings show that, as with exploit kits, the complexity of client-side phishing code yields dynamic traces that can track phishers' kit deployments in the wild, enabling automatic attribution of multi-domain and multi-lingual deployments.

%% file: tex/appendix.tex
\appendix
\section*{Ethics}
\label{sec:ethics}
This research relied on publicly and commercially available URLs detected using proprietary methods to be phishing websites. All of the crawling traffic originated from a network designated for web measurement, with the researchers monitoring the abuse contact for that IP space. We did not collect or store any identifiable information about the individuals behind phishing kits or the pages, and we did not conduct any live testing of their infrastructure. The phishing kits, on the other hand, contain identifiable information about their victims and Telegram API keys to access exfiltration group chats, and can be trivially manipulated to bypass current signature-based detection tools and redeployed in the wild. For this reason, we require a data-sharing agreement before we share the kits collected with any future researchers.
\section*{Appendix A}
\input{figures/ms_defender.tex}
\input{figures/brand_vs_phishing.tex}
\input{figures/timeline.tex}
\input{figures/AllTechniques.tex}
\input{figures/kit_binding.tex}    %

%% file: figures/ms_defender.tex
\begin{figure*}[ht]
  \centering
  \subfloat[\centering Page variant in English]{\fcolorbox{black}{white}{\includegraphics[width=3in]{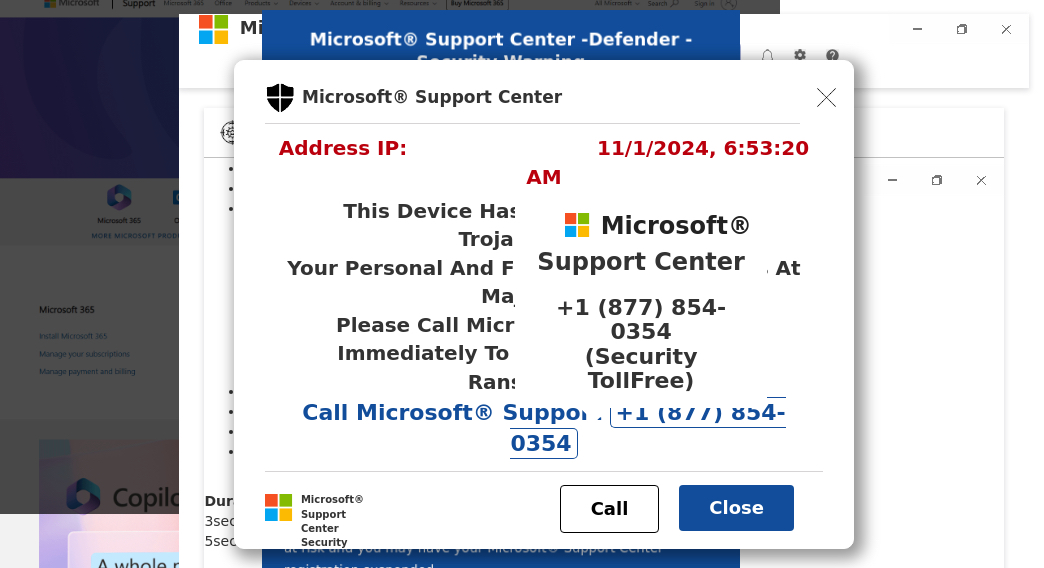} }}%
  \qquad
  \subfloat[\centering Page variant in Japanese]{\fcolorbox{black}{white}{\includegraphics[width=3in]{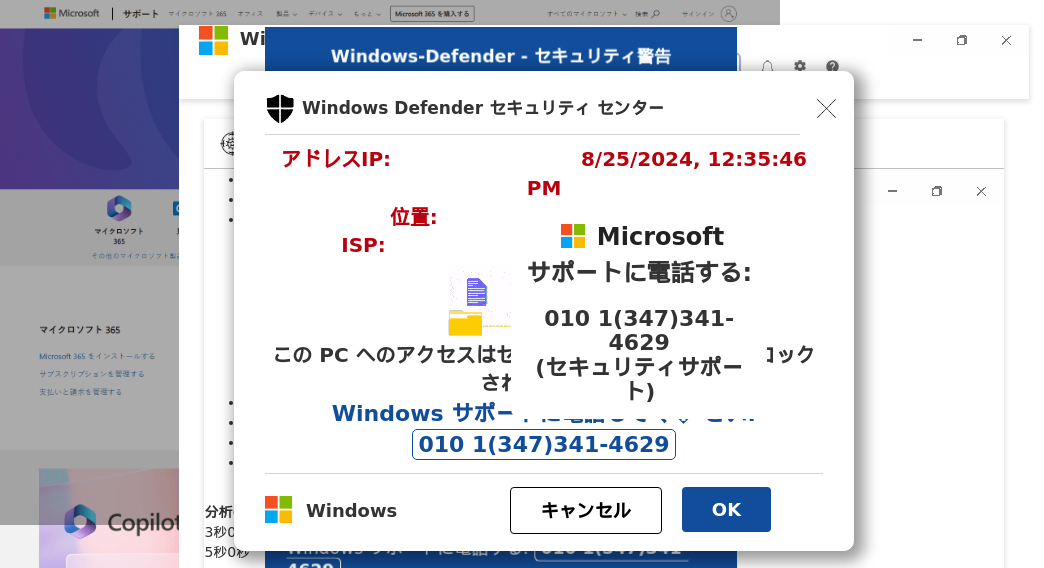} }}%
  \caption{Cropped screenshots from Cluster-98b0e6bb, IP addresses and location redacted to ensure anonymity.}%
  \label{fig:ms_defender}%
\end{figure*}

%% file: figures/brand_vs_phishing.tex
\begin{table*}[ht]
\caption{Comparison of browser APIs executed by phishing pages against the original login page they target}
  \centering
\begin{tabular}{l|r|r|r|r|r|r|r}
\textbf{Brand}  & \textbf{Phishing pages observed} & \textbf{Avg \% Similar} & \textbf{Std Dev (\%)} & \textbf{Perfect Subset} & \textbf{50\%+ Match} \\
\hline
Facebook &  338,686 & 11.6 & 9.6 &  0 & 24,039 \\
USPS &  66,692 & 10.3 & 10.5 &  0 & 53 \\
Meta &  51,969 & 12.4 & 7 &  0 & 2,605 \\
Microsoft & 3,768 & 11.8 & 8.7 &  0 & 419 \\
IRS & 1,207 & 11.5 & 9 &  0 & 61 \\
\end{tabular}
\label{tab:phishing-metrics}
\end{table*}

%% file: figures/timeline.tex
\begin{figure*}[t]
  \includegraphics[width=\textwidth]{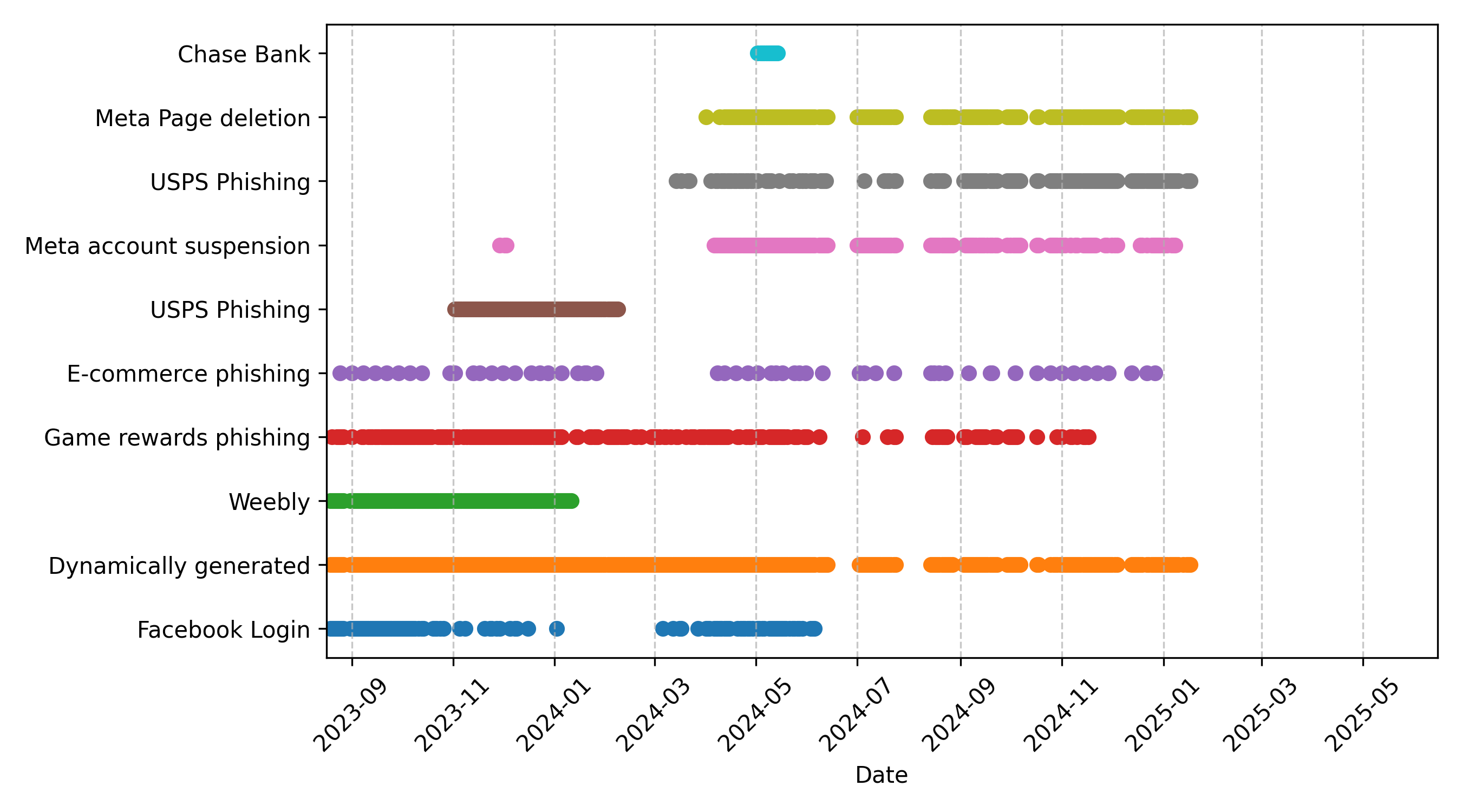}
  \caption{Timeline of the top 10 clusters based on the number of pages. We note that for eight out of the ten clusters, we see the clusters persistently re-appear in the ecosystem.}
  \label{fig:top_10_clusters}
\end{figure*}

%% file: figures/AllTechniques.tex
\begin{figure*}[t]
  \centering{
  \includegraphics[width=\textwidth]{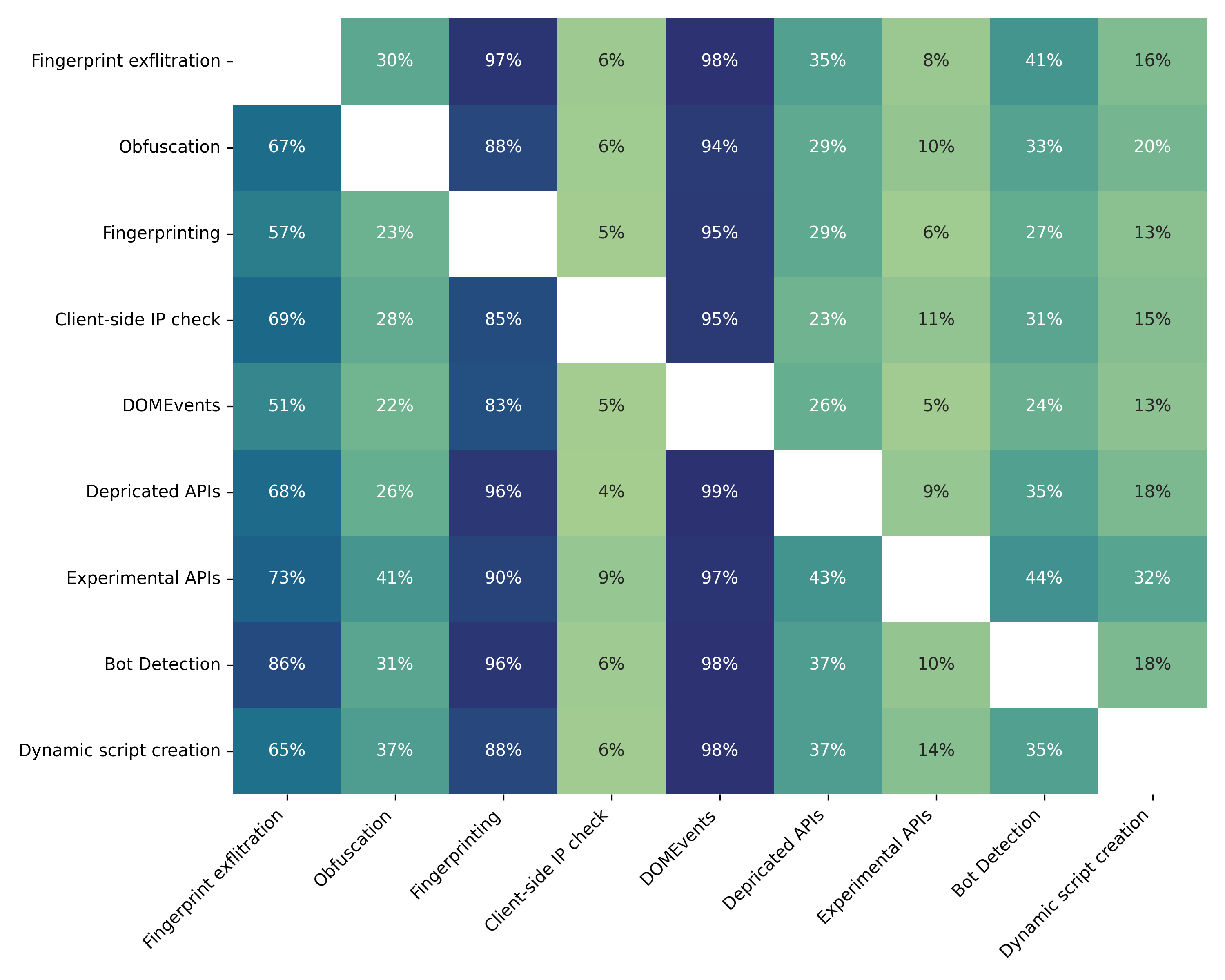}
  \caption{Confusion matrix between all of the techniques enumerated and cluster lifetime characteristics.}
  \label{fig:AllTechniques}
  }
\end{figure*}

%% file: figures/kit_binding.tex
\begin{figure*}[ht]
  \includegraphics[width=\textwidth]{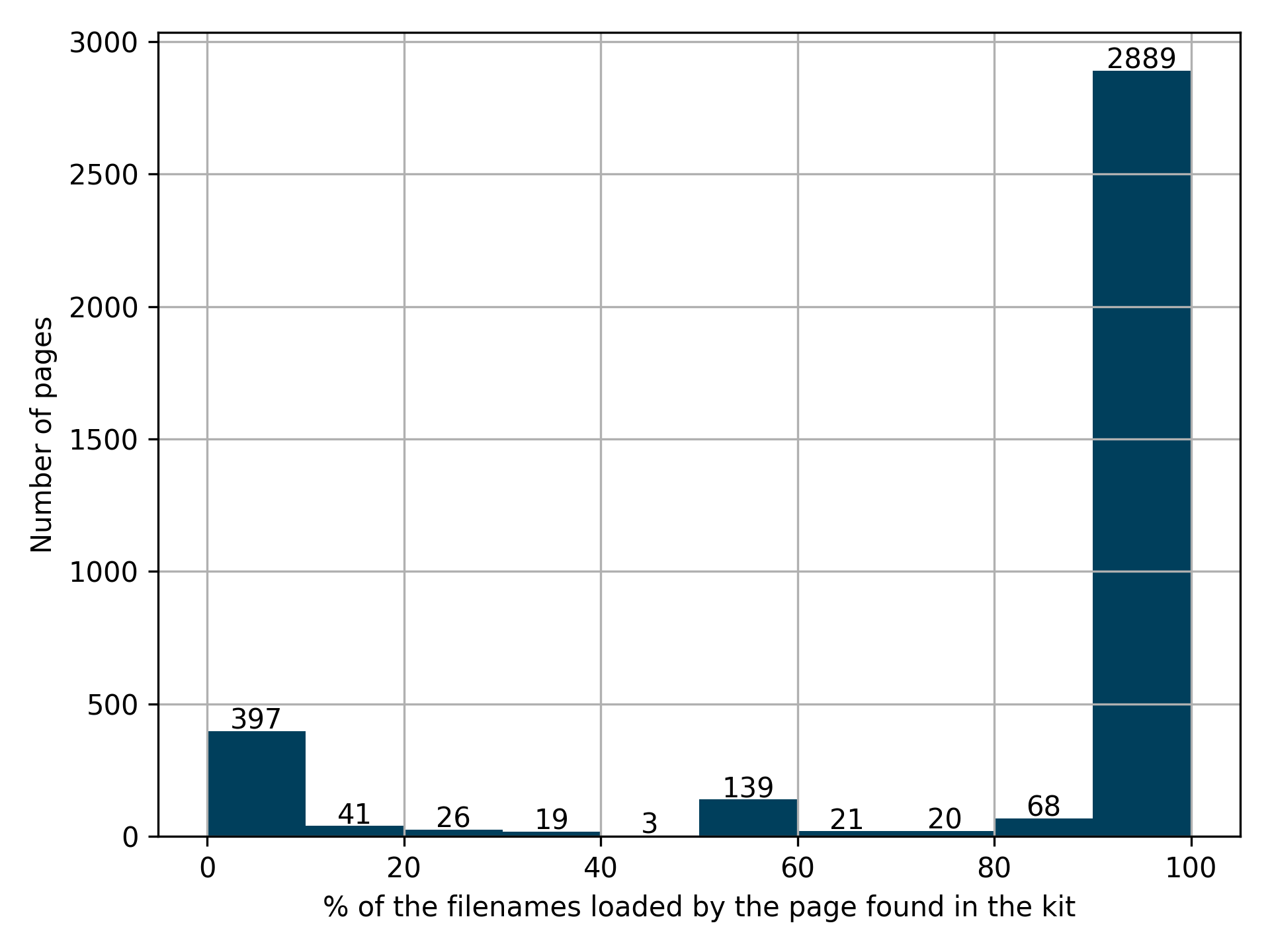}
  \caption{Histogram distribution of the percentage match between the filenames loaded by the website in a first-party context (images, stylesheets, JavaScript code) and filenames in the extracted zipfiles. We note that a 0\% match means the binding is inconclusive, as kits that load all assets inline would fail to bind.}
  \label{fig:kitbindbing}
\end{figure*}